\documentclass[12pt]{iopart}
\usepackage{graphicx,iopams,amssymb,mathrsfs,eufrak,color}
\begin{document}

\title[Vacancy-induced localized modes and impurity band formation . . . ]
{Vacancy-induced localized modes and impurity band formation in the
Haldane model: a quantum dot analogy}
\author{Hussein Alshuwaili$^a$, Zahra Noorinejad$^b$, Mohsen Amini$^a$, Morteza Soltani$^a$, Ebrahim Ghanbari-Adivi$^{a*}$ }
\address{$a)$ Faculty of Physics, University of Isfahan, Isfahan 81746-73441, Iran}
\address{$b)$ Department of Physics,  Islamic Azad University - Shahreza Branch~(IAUSH), Shahreza, Iran}
\ead{ghanbari@phys.ui.ac.ir} \vspace{10pt}
\date{}
\begin{abstract}
In this study, the Haldane model's edge states are utilized to
illustrate that a zero-energy localized state forms around a single
vacancy in the model. In order to complete this task, the
conventional unit cell associated to the Haldane hexagonal structure
is transferred onto a two-leg ladder in momentum space, effectively
forming an extended Su-Schrieffer-Heeger~(SSH) lattice through a
one-dimensional Fourier transform. Through the application of a
suitable unitary transformation, the two-leg SSH ladder in momentum
space is converted into an equivalent lattice with two distinct
on-site states with different momentum that are suitable for the
calculations. Ultimately, the desired zero-energy localized mode
formed around the vacant-site is represented by a combination of the
armchair edge states. Furthermore, the scenario involving two vacant
sites is investigated and it is revealed that an effective hopping
interaction exists between the localized states formed around the
on-site vacancies created along a zigzag chain in the lattice. This
structure can be likened to the structure of a quantum dot with two
none-degenerate energy levels. Such a hopping interaction is absent
for the same vacancies  created on the armchair chains. Finally, it
is shown that introducing vacancies periodically on the sites of a
zigzag row along a finite-width ribbon with the Haldane structure
leads to the emergence of an impurity band within the energy gap.
\end{abstract}
\section{Introduction\label{Sec01}}
Two-dimensional~(2D)~materials are a class of nanostructured
low-dimensional materials with unique intresting properties that
usually consists of a single layer of atoms and are often well
suited for the applications where the bulk material would be
unsuitable~\cite{Novoselov01,Bondavalli01,Arul01,Kumbhakar01}. The
laboratory production of graphene as the first well-known 2D
material isolated in 2004, as well as the growth of the applications
of the 2D~materials in industry and technology, has sparked
considerable interest in studying the properties of such
structures.\par
Transition metal dichalcogenides~(TMDs) are a class of the
2D~materials composed of a transition metal combined with one of the
chalcogen elements~\cite{Arul01}. These compounds exhibit a layered
structure, similar to graphene, with strong covalent bonding within
the layers and weak van~der~Waals interactions between the layers.
This structure gives rise to unique properties, including high
carrier mobility, sizable band gaps, and strong light-matter
interactions. As a result, TMDs are of significant importance in
various applications such as electronics, optoelectronics and energy
storage. They have been explored for use in transistors,
photodetectors and solar cells due to their exceptional electrical
and optical properties~\cite{Arul01}.\par
The study of the electronic structure of 2D~materials under the
influence of edge and/or localized modes within these structures is
among the top interests for researchers.  These states can play a
crucial role in the optical absorption and carrier transport
properties of 2D~materials. In addition to these facts, exploring
the behavior of localized states is also interesting in
investigating the properties of the disordered systems, and it can
be used in explaining the transport properties of materials such as
graphene and phosphorene as a single layer of black phosphorus.
Exploring the impact of vacancies has further heightened interest in
this subject, with numerous studies focusing on the behavior of
these states around vacant sites in 2D~materials. For instance,
research in~\cite{Pereira01,Pereira02,Sadeghizadeh01} demonstrates
that the creation of a vacant site in a single-layer graphene sheet
results in the formation of a localized state around the vacancy.
Other researches have looked at the same issue for a bilayer
graphene and and a monolayer phosphorene
sheet~\cite{Amini01,Castro01}.
It is also known that if the vacancies are periodically placed in
the system, they alter the electronic band structure of the system
and lead to the formation of new energy bands. These new bands are
known as impurity bands. For example, it has been shown in
reference~\cite{Rezaei01} that the existence of a periodic set of
vacancies in an armchair phosphorene nano-ribbon leads to the
creation of an impurity band in the energy gap of the system. The
advantage of such systems is that the width of the impurity band can
be adjusted by alerting the distance between the vacant sites, and
this can be applied to improve the design of thermoelectric systems.
\par
Also, prediction and explanation of the topological insulating
behavior of the 2D~materials in general and TMDs in particular is
one of the topics of great interest to researchers~\cite{Asboth01}.
The localized states in 2D~materials and their topological
insulating behavior are interconnected through the electronic
structure and topology of these materials. Understanding the
relationship between these factors can help researchers develop new
materials and explore unique electronic properties in 2D~systems. On
the other hand, the presence of the impurity bands can be used to
develop the construction and application of the topological
insulators. In recent years, this fact has led to an intense focus
on 2D~materials for the construction and design of topological
insulators.\par
In 1988, Haldane introduced the first 2D~topological structure that
relied on the graphene model~\cite{Haldane01}. The Haldane model
describes a theoretical approach to understanding the electronic
properties of graphene in the presence of a broken time-reversal
symmetry, which can occur due to factors such as magnetic field or
substrate interactions. This model introduces next-nearest-neighbor
complex hoppings in the graphene lattice, leading to a non-trivial
band structure with a non-zero Chern number, resulting in a quantum
Hall effect and topologically protected edge states. This model has
been influential in studying topological phases and exotic
electronic behaviors in graphene and related materials.\par
By incorporating the spin-orbit interaction into the Haldane's
approach, Kane and Mele  were able to explain the properties of the
2D~$Z2$ topological insulators~\cite{Kane01}. This in fact provides
a theoretical framework for understanding the electronic properties
of graphene, taking into account intrinsic spin-orbit coupling.\par
After several years of the above mentioned pioneering works by
Haldane, Kane and Mele, the discovery of moir\'{e} patterns in
2D~layered materials and the possibility of making Haldane and $Z2$
topological insulators using the moir\'{e} superlattices, the issue
of 2D~topological insulators has been brought to attention
again~\cite{He01}. Moir\'{e} patterns are interference patterns that
occur when two or more layers of a 2D~crystal are stacked with a
slight twist angle~\cite{Xiao01}. The relations between TMDs,
moir\'{e} patterns in 2D~layered materials and  the Haldane model
has been investigated by MacDonald and his co-workers~\cite{Wu01}.
In their outstanding work, it has been demonstrated that the
moir\'{e} patterns formed by a bilayer structure of TMDs can exhibit
topological bands with non-zero Chern number. Investigation of the
effects of vacancies on the futures of topological insulators is a
current topic of interest in topological structures. Due to the
mentioned interest, many experimental and numerical studies have
been conducted on the
topic~\cite{Miranda01,Lima01,Loio01,Xu01,Meier01,Li01,Orth01,Assaad01,Stuhler01}.
For instance, the properties of localized states surrounding vacant
sites in the 2D~topological systems have been numerically examined
in references~\cite{Miranda01} and~\cite{Lima01}. Moreover, as
reported in~\cite{Stuhler01}, numerical and experimental analysis
have been carried out to examine the effects of a vacant region in a
topological insulator.\par
In the present study, the effects of the presence of the vacant
sites in 2D~Haldane materials has been investigated. First, the wave
function of the created localized state around a single vacancy in
the Haldane lattice has been obtained. For this purpose, by using
appropriate Fourier and unitary transformations, the unit cell of
the crystal is mapped onto a two-ladder~SSH~ladder in the momentum
space. The final deduced SSH~structure is suitable for calculating
the localized modes in terms of the armchair edge states. By
combining the on-site states with different energies in this
structure,  the localized state surrounding the empty lattice site
is acquired. The structure of the obtained localized modes is
similar to that of a single-level quantum dot. In the following, the
hopping interaction energy between the localized modes formed around
two vacant sites created on a line along the armchair and/or zigzag
directions is studied. It is discussed that this hopping interaction
does not exist for vacancies positioned along the armchair
direction, while it remains notable for the vacancies situated in
the zigzag direction even over relatively considerable distances.
For the latter case, the interacting behavior of the two zero-energy
localized states surrounding the vacant sites resembles that of a
two-state quantum dot. At the end, by considering the vacant sites
that are placed periodically on a chain in the zigzag direction of a
hexagonal structure, the characteristics of the impurity band
created in the electronic band structure of the system have been
investigated.\par
The outline of the rest of the article is as follows. In the next
section, we will study the presence of a single vacancy in the
Haldane model and construct the zero-energy localized state created
around the vacancy. In the next section, the number of the vacant
sites is increased to two and it is shown that for the vacancies
created sufficiently close to each other on a zigzag chain of the
lattice, there exists a non-zero hopping interaction energy between
the localized states generated around the vacant sites. In this
section, it is also discussed that if the number of the vacant sites
increases and they are placed periodically on a lattice zigzag
chain, the impurity bands will appear in the energy gap. Finally,
the conclusion remarks are provided in the last section.
\begin{figure}[t]
\begin{center}
\includegraphics[scale=0.3]{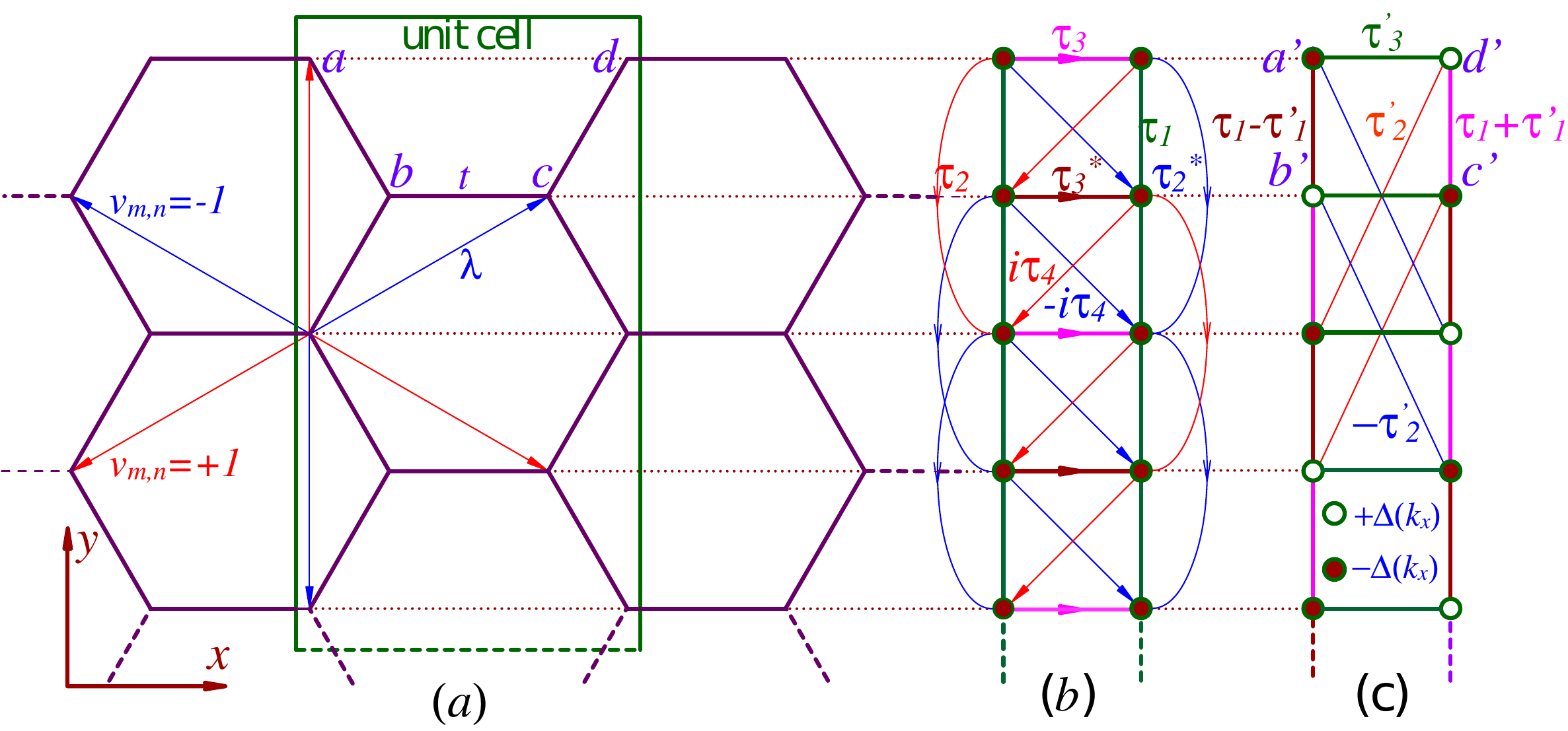}

\end{center}
\caption{(a) Haldane hopping parameters of a semi-infinite honeycomb
lattice with an armchair side including a conventional unit cell
comprises of four distinct atoms of $a$, $b$, $c$, and $d$.  (b) The
Hamiltonian associated with the conventional unit cell of the
considered structure in panel (a) is mapped onto a symmetric two-leg
SSH ladder in the momentum space with some complex hopping
parameters denoted on the graph by using a Fourier transform along
the $x$ axis followed by a proper gage transformation, (c) A proper
unitary transformation maps the  symmetric extended SSH structure
displayed in panel~(b) to an asymmetric two-leg SSH ladder including
two distinct sublattices with on-side energies of $\pm \Delta(k_x)$
where $\Delta(k_x) = t \sin (k_x/2)$. \label{Fig01}}
\end{figure}
\section{Calculation of the localized state around a single vacancy\label{Sec02}}
As is mentioned in the previous section, TMDs are a class of
2D~layered materials that exhibit unique electronic properties due
to their hexagonal lattice structure~\cite{Arul01,Xiao01}. When TMDs
are placed on a substrate with a slightly different lattice
constant, moir\'{e} patterns can form due to the interference
between the two lattices~\cite{He01,Xiao01}. These moir\'{e}
patterns can have a significant impact on the electronic structure
of the TMDs, leading to the emergence of unconventional quantum
phenomena such as correlated insulating states, superconductivity,
and topological phases. Recently, the Haldane model, which describes
the behavior of electrons in a honeycomb lattice with broken
inversion symmetry and spin-orbit coupling, has been used to study
the effects of moir\'{e} patterns on the electronic properties of
TMDs, providing insights into the emergence of novel quantum states
in these materials. One of our primary motivations for current study
is investigation on the topological properties of moir\'{e} bands in
a twisted bilayer of TMDs in presence of the on-site vacancies.
Therefore, we focus on the study of the effect of the on-site
vacancies on the electronic structure of the hexagonal lattices in
the Haldane model. The structure of interest can be a graphene or
phosphorene sheet  or a superlattice created by moir\'{e} patterns.
With this aim, this section involves the calculation of the
localized state surrounding a single vacancy within the Haldane
model. The followed approach can be used also for Kane-Mele model
due to the fact that in presence of the axial spin symmetry this
model is equivalent to two decoupled copies of Haldane model for
electrons in opsin spin directions.\par
Introducing an on-site vacancy on a hexagonal lattice sheet divides
it into two half sheets with armchair (or zigzag) edges. The
armchair edge modes spanned the entire first Brillouin zone within
the range of $[-\pi, +\pi]$, whereas the zigzag edge modes only
extend across the interval of $[\pi/3, 2\pi/3]$ of this
zone~\cite{Rahmati01}. Therefore, we first focus on examining the
edge modes of a hexagonal half-sheet with an armchair edge.
Panel~(a) in figure~\ref{Fig01} illustrates a schematic diagram of
the specific system which will be discussed in this step. As
depicted in the figure, the system of interest comprises a
2D~honeycomb lattice with a Haldane Hamiltonian. Explicitly, this
Hamiltonian can be expressed in the following form~\cite{Haldane01}:
\begin{equation}\label{Eq01}
{\cal H}= t \sum\limits_{\langle m,n\rangle}^{}  c_m^\dagger c_n +
\lambda \sum\limits_{\langle\langle m,n\rangle\rangle}^{}
e^{i\nu_{mn}\varphi} c_m^\dagger c_n ,
\end{equation}
in which ${\langle m,n\rangle}$ and $\langle\langle
m,n\rangle\rangle$ demonstrate respectively the nearest-neighbor and
the next-nearest-neighbor pairs of sites, while $t$ and $\lambda$
are their corresponding hopping energies. $c_m$ and $c_m^\dagger$
are the on-site electron annihilation and creation operators,
respectively. $e^{i\nu_{mn}\varphi}$ is a phase factor in which
$\nu_{mn}$ takes $\pm 1$ depending on the positions of sites $m$ and
$n$ relative to each other in the lattice. Since the mass term, a
staggered on-sublattice potential with a given strength, does not
play an important role in the following discussions, it has been
excluded from the Haldane Hamiltonian given in
equation~\ref{Eq01}.\par
To construct the localized modes of the specified system, we are
looking for the zero-energy eigenstates of the Hamiltonian  given in
equation~\ref{Eq01}.Various techniques are available for achieving
the localized states in 2D~structures. For instance, in
reference~\cite{Pereira02}, using a theorem in linear algebra, it
has been demonstrated  that the unequal distribution of vacant sites
between the two sublattices will result in the appearance of
zero-energy localized modes. In fact, if the Hamiltonian of a system
with sublattices $A$ and $B$ can be represented as
\begin{equation}\label{Eq02}
\cal{H} = \left[
\begin{array}{cc}
\epsilon_A {\bold 1}_A     &     H_{AB}\\
H_{AB}^\dagger             &   \epsilon_B {\bold 1}_B
\end{array} \right] ,
\end{equation}
then the presence of a vacancy in subspace $A$ will result in the
emergence of a zero-energy localized state denoted as
$\left|\psi_B^L\right\rangle$ in subspace $B$ which satisfies the
eigenvalue equation of,
\begin{equation}\label{Eq03}
{\cal H} \left|\psi_B^L\right\rangle = 0.
\end{equation}
In equation~\ref{Eq02}, ${\bold 1}_A$ and ${\bold 1}_B$ are the
identity matrices in subspaces $A$ and $B$, $\epsilon_A$ and
$\epsilon_B$ are local energies on the sublattices and $H_{AB}$ is
the $n_A\times n_B$ hopping matrix with ${n }_A$ and ${n}_B$ as the
dimensions of the subspaces. In summary, the theorem indicates that
a single vacancy in the sublattice $A$ introduces a localized zero
mode in the other sublattice denoted by $B$. However, the mentioned
theorem is not particularly applicable to the Haldane model. This is
primarily due to the lack of clarity in converting the Haldane
Hamiltonian from the form given in equation~\ref{Eq01} to the form
of equation~\ref{Eq02}, and also due to the challenging nature of
obtaining the analytical form of the localized state using this
approach.\par
To overcome the difficulties mentioned above, we use another method
in our current study. This approach involves making the proper
Fourier and unitary transformations to reduce the system's
Hamiltonian to a proper two-leg extended SSH ladder and constructing
the zero-energy localized states by a superposition of armchair edge
states. This technique has been implemented in the earlier studies
to determine the localized states of both graphene and phosphorene
in references~\cite{Pereira01,Pereira02} and~\cite{Amini01},
respectively. Since the closed form of the topological armchair edge
states for the Haldane's graphene model have been derived previously
in reference~\cite{Sadeghizadeh01}, the adopted approach seems to be
effective. In fact, the accessibility of these closed forms
streamlines our efforts to derive the representation of the
localized states surrounding the single vacant site in the Haldane
model in terms of these states.\par
\begin{figure}[t]
\begin{center}
\includegraphics[scale=.3]{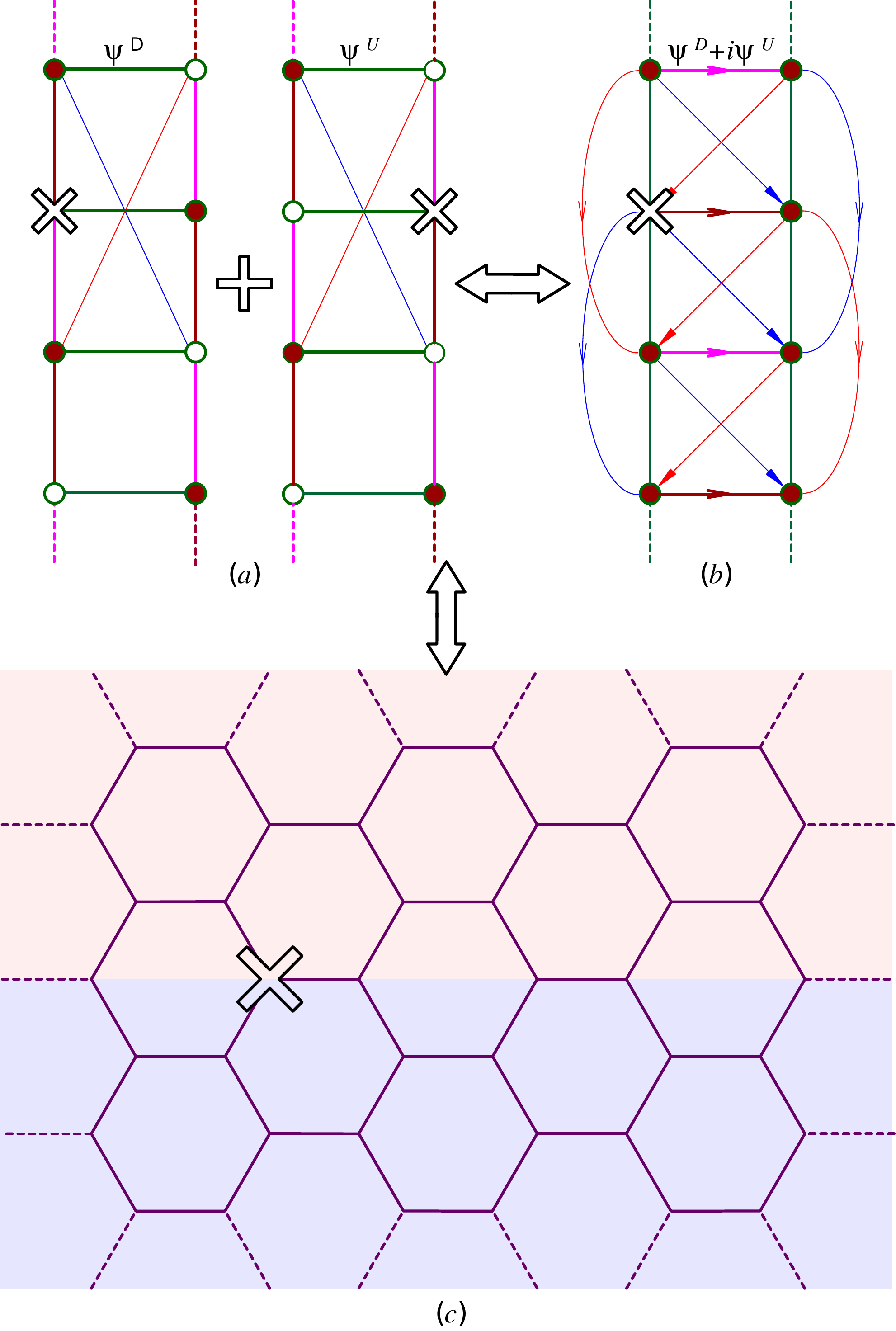}

\end{center}
\caption{(a) Placing the vacancy on site $b'$ leads to an edge state
which is complectly localized in the lower part of the
asymmetric~SSH~ladder, whereas placing it on site $c'$ result in a
completely localized edge state in the upper part of that, (b) The
linear combination of the two localized edge states in the upper and
lower parts of the asymmetric two-leg~SSH~ladder creates a localized
state on site $b$ in the symmetric extended~SSH  lattice in momentum
space. (c) With the inverse Fourier transform, the structure of the
localized state around site~$b$ in momentum space is mapped to the
localized state in a sheet of the hexagonal structure. As can be
seen, the vacancy divides the sheet into two semi-finite ribbons
with armchair sides. \label{Fig02}}
\end{figure}
As demonstrated in figure~\ref{Fig01}, taking $x$ and $y$ axes along
the armchair and zigzag chains of the lattice, each site can be
labeled by a pare of $(\ell, s)$, where $\ell$ and $s$ are zigzag
and armchair chain numbers. Let us consider the conventional unit
cell of the lattice comprises four distinct sites labeled as $a$,
$b$, $c$ and $d$  as displayed in the first panel of
figure~\ref{Fig01}. The figure illustrates that the system is
invariant under translation in the $x$ direction, which means that
Bloch wavenumber $k_x$ is a good quantum number. This fact helps us
to map the Hamiltonian to a two-leg SSH ladder in the momentum space
as shown in panel~$(b)$ of figure~\ref{Fig01}. This is acquired
using the Fourier transformation of the electron annihilation and
creation operators of the four bases of the unit cell as
\begin{equation}\label{Eq04}
\big(\begin{array}{l}
a_{\ell,k_x}\\
b_{\ell,k_x}\end{array}\big) =\sum\limits_{s}^{}
\big(\begin{array}{l}
a_{\ell,s}\\
b_{\ell,s}\end{array}\big) e^{ik_x s},\qquad \big(\begin{array}{l}
c_{\ell,k_x}\\
d_{\ell,k_x}\end{array}\big) =\sum\limits_{s}^{}
\big(\begin{array}{l}
c_{\ell,s}\\
d_{\ell,s}\end{array}\big) e^{ik_x (s+{1\over 2})}.
\end{equation}
To maintain the simplicity of the hopping parameters in the extended
SSH~lattice, the extra phase appearing in the above transformation
for $c_{\ell,s}$ and $d_{\ell,s}$ operators is generated using a
proper gage transformation. The hopping parameters of the converted
lattice are $\tau_1 = t,\quad \tau_2=\lambda,\quad \tau_3=t
e^{ik_x/2}, \tau_4 = -2 i \lambda \cos(k_x/2)$ where as is some of
them are complex. As can be seen from panel~(b) of
figure~\ref{Fig01}, the structure created by these transformations
is symmetric, so it is impossible to determine a subspace in which
the wave function of the edge mode is zero. To solve this issue, by
applying a proper unitary transformation given by
\begin{equation}\label{Eq05}
\Big[\begin{array}{l}
a'_{\ell,k_x}\\
d'_{\ell,k_x}\end{array}\Big] ={\sqrt{2}\over 2}
\Big[\begin{array}{ll}
1 & i\\
i & 1 \end{array}\Big]\Big[\begin{array}{l}
a_{\ell,k_x}\\
d_{\ell,k_x}\end{array}\Big] ,\quad \Big[\begin{array}{l}
b'_{\ell,k_x}\\
c'_{\ell,k_x}\end{array}\Big] ={\sqrt{2}\over 2}
\Big[\begin{array}{ll}
1 & i\\
i & 1 \end{array}\Big]\Big[\begin{array}{l}
b_{\ell,k_x}\\
c_{\ell,k_x}\end{array}\Big],
\end{equation}
this symmetric structure can be mapped into a similar asymmetric
structure that includes two separate sublattices. As is seen from
panel~(c) in figure~\ref{Fig01}, the different on-site energies for
atoms in two sublattices are denoted as $+\Delta (k_x)$ and $-\Delta
(k_x)$, respectively.\par
The hopping parameters of the asymmetric SSH~structure are given in
panel~(c) of figure~\ref{Fig01} in which  $\tau'_1 = -2 \lambda
\cos(kx/2),\ \tau'_2 = \lambda,\ {\rm and}\ \tau'_3 = t
\exp{(ik_x/2)}$. Also, the Hamiltonian corresponding to this
structure can be considered as the sum of two separate parts, which
is given by
\begin{equation}\label{Eq06}
{\cal H}' = {\cal H}'_0 + {\cal H}'_1.
\end{equation}
${\cal H}'_0$  is the hopping Hamiltonian that gives the zero-energy
flat-band edge states and ${\cal H}'_1$ is the on-site Hamiltonian
that acts as a perturbing Hamiltonian and is responsible for the
edge state dispersion.\par
As previously stated, the armchair edge mode is an eigenmode of the
whole Hamiltonian ${\cal H}'_0$. Clearly, the on-site Hamiltonian,
${\cal H}'_1$ has no any impact on such an eigemode. Thus, in order
to construct the intended edge mode, we must simply discover the
eigenmode of the hopping Hamiltonian, ${\cal H}_0$. Such an eigemode
satisfies  the eigenvalue Schr\"{o}dinger equation of
\begin{equation}\label{Eq07}
{\cal H}'_0\big|\psi_e(k_x)\big\rangle = 0.
\end{equation}
The wave function associated with the state that satisfies the above
equation must be non-zero in one sublattice while vanishes in the
other. By referring to the asymmetric SSH~structure shown in
panel~(c) of figure~\ref{Fig01}, and examining the hopping
parameters appearing in this structure, we guess that the desired
wave function on sites $a'_{\ell,k_x}$ and $c'_{\ell,k_x}$ has a
non-zero value but vanishes on $b'_{\ell,k_x}$ and $d'_{\ell,k_x}$.
With such an assumption, we can assume that the desired edge state
is a linear combination of the bases $\big|a'_i (k_x)\big\rangle$
and $\big|c'_i(k_x)\big\rangle$ in the form of
\begin{equation}\label{Eq08}
\big|\psi_e(k_x)\big\rangle = \sum_{i = 1} q_{2i-1}
\big|a'_i(k_x)\big\rangle + q_{2i} \big|c'_i(k_x)\big\rangle,
\end{equation}
where the expansion coefficients, $q_{2i-1}$ and $q_{2i}$, represent
the relative amplitudes of the different component waves of
$\big|a'_i(k_x)\big\rangle$ and $\big|c'_i(k_x)\big\rangle$ in the
wave function. By substituting the proposed ket state~\ref{Eq07} in
the eigenvalue equation of~\ref{Eq08}, and choosing the initial
coefficients $q_1$ and $q_2$, some recursive relations can be
derived that using them and conducting numerical computations allows
for the unique determination of various coefficients based on the
lattice's hopping parameters.\par
Since a closed form for the expansion coefficients appearing in
equation~\ref{Eq05} is not accessible, we are unable to provide an
analytical form for the armchair edge mode being studied. Anyway,
the numerically obtained ketstate is also the eigenket of ${\cal
H}'_1$, so that satisfies the eigenvalue equation of
\begin{equation}\label{Eq09}
{\cal H}'_1\big|\psi_e(k_x)\big\rangle = t \sin(k_x/2)
\big|\psi_e(k_x)\big\rangle,
\end{equation}
indicating that the on-site energies of the lattice are given by
$\pm \Delta(k_x) = \pm t \sin(k_x/2)$.\par
Now, we are going to utilize the approach outlined in
references~\cite{Pereira01,Amini01} and~\cite{Castro01} to derive
the localized state surrounding a single on-site vacancy in the
Haldane lattice through a superposition of the obtained armchair
edge states within the range of $[-\pi,+\pi]$. Considering
figure~\ref{Fig02} assists us to achieve this goal. As depicted in
the figure, we assume that the vacancy is positioned at the place of
$b'$ on an infinite asymmetric two-leg SSH~ladder within the space
of $a',\ b',\ c',{\rm and}\ d'$. Applying the same procedure
employed in reference~\cite{Sadeghizadeh01} demonstrates that a
localized state emerges in the lower half of the ladder around the
vacancy. This state, denoted as $|\psi^D_e(k_x)\rangle$, is entirely
absent in the upper half of the ladder. If, as depicted in panel~(a)
of figure~\ref{Fig02}, the vacancy is situated at site of $c'(k_x)$,
the localized state formed around the vacancy will be confined to
the upper half of the ladder, thereby being entirely missing from
the lower half. In the following, we refer to this localized state
by $|\psi^U_e(k_x)\rangle$. Applying the inverse of the
transformation given in equation~\ref{Eq05}, we come to conclude
that the localized state for the extended SSH~structure dipected in
panel~(b) of figure~\ref{Fig02}, is a combination of
$|\psi^D_e(k_x)\rangle$ and $|\psi^U_e(k_x)\rangle$, so that, for
each arbitrary value of $k_x$ within the interval of
$\left[-\pi,+\pi\right]$, we have:
\begin{equation}\label{Eq10}
|\psi_e (k_x)\rangle = {1 \over \sqrt{2}} \Big(
|\psi^D_e(k_x)\rangle + i |\psi^U_e(k_x)\rangle \Big).
\end{equation}
If the vacancy is regarded on site $d(k_x)$, the corresponding edge
state is formed with a similar superposition in which $i$ is changed
to $-i$. This general state can be a linear combination of the
eigenstates associated with these eigenvalues.\par
Without loss of generality, we assume that the vacancy is fixed on
origin site of $(\ell,s)=(0,0)$.  To acquire the localized state
surrounding the vacancy, it is necessary to combine the states of
$|\psi_e(k_x)\rangle$, found in equation~\ref{Eq08}, with each other
in such a way that the condition of
\begin{equation}\label{Eq11}
\langle\ell, \pm 1|\psi_e(k_x)\rangle = {\Big(
\begin{array}{c} 1\\ i\end{array}\Big)}C(k_x) e^{ik_x\ell},
\end{equation}
is satisfied for all sites located in the row below or above the
empty site. On the second side of the above equation,  $C(k_x)$ and
$i C(k_x)$ are the amplitudes of the wave function in the rows above
and below the vacant site, respectively.  In fact, during the
superposition of the states of $|\psi_e(k_x)\rangle$ to making the
zero-energy localized mode, this condition has been taken into
account instead of the normalization condition. According to this
equation and using the identity of
\begin{equation}\label{Eq12}
\int_{-\pi}^{+\pi} e^{ik\ell} dk = 2\pi \delta_{\ell,0},
\end{equation}
we come to conclude that the localized state of $|\psi^L\rangle$ can
be represented as a superposition of the states of
$|\psi_e(k_x)\rangle$ for all possible values $k_x$ in the first
Brillouin zone. Such a superposition reads
\begin{equation}\label{Eq13}
|\psi^L\rangle=N \int_{\pi}^{+\pi} {\left|\psi_e(k_x)\right\rangle
\over C(k_x)} dk_x,
\end{equation}
where $N$ is a normalization factor. Of course, it should be noted
that since the  atoms $d_{\ell,k_x}$ follow the Fourier transform
of~\ref{Eq04}, the wave function is non-zero on this type of atoms
located in the first row below the vacancy adjacent to atoms
$a_{\ell,k_x}$. Also, it is important to mention that, on the row
that the vacancy has been created, since both wave functions of
$|\psi^D_e(k_x)\rangle$ and $|\psi^U_e(k_x)\rangle$ are non-zero,
the wave function obtained from their superposition is non-zero
too.\par
It is remarkable to notify that the obtained wavefunction satisfies
the eigenvalue equations of
\begin{equation}\label{Eq14}
{\cal H}'_0|\psi^L\rangle = 0,\qquad {\rm and}\qquad {\cal
H}'_1|\psi^L\rangle = 0.
\end{equation}
Also, since the perturbing Hamiltonian ${\cal H}'_1$ is an odd
function of momentum, $k_x$, its expectation value on the obtained
zero-energy localized state is zero, $\left\langle \psi_L\right|
{\cal H}'_1 \left| \psi_L\right\rangle = 0$.\par
\begin{figure}[t]
\begin{center}
\includegraphics[scale=0.6]{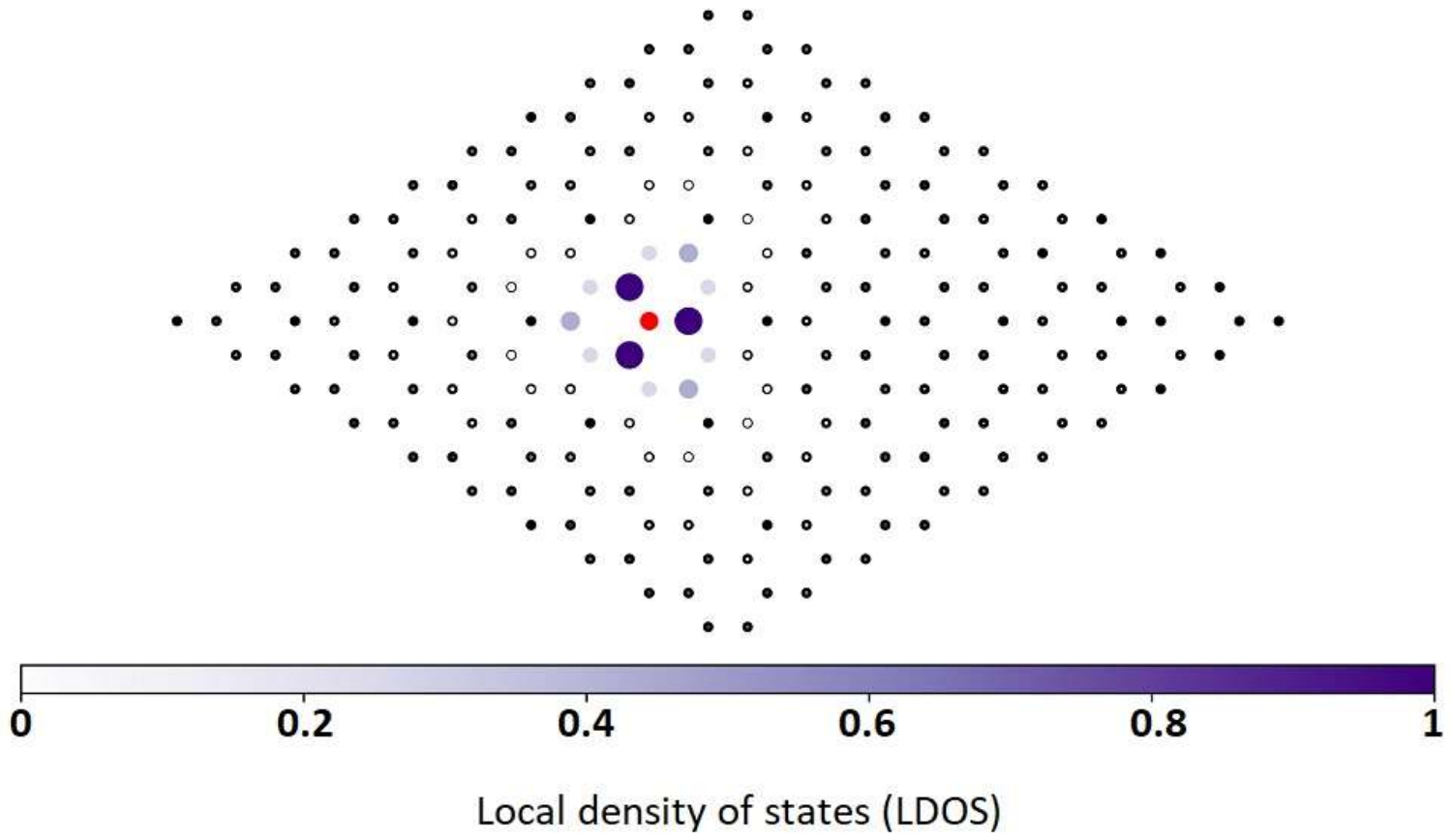}

\end{center}
\vspace*{-3mm} \caption{Zero-energy local density of states in a
sheet of  Haldan hexagonal  structure with a single on-site vacancy
in which the vacant site is shown in red \label{Fig03}}
\end{figure}
As mentioned before, the closed forms of edge states
$\left|\psi_e(k_x)\right\rangle$, as defined in equation~\ref{Eq08},
are not reachable. Due to this limitation, we cannot derive a closed
form for the localized states under investigation. However, in order
to validate the current model, we aim to assess the accuracy of its
predictions by comparing them with the outcomes of exact numerical
computations. For this purpose, a graphical representation of the
local density of states~(LDOS) produced through exact numerical
computations is  displayed in figure~\ref{Fig03}. Since the
numerical results obtained in this research are nearly identical to
those displayed in the figure, we have avoided representing. All the
numerical calculations depicted in this figure have been performed
using the Pyqula library~\cite{Pyqula01}. As can be seen from the
figure, the wave function above and below the armchair row in which
the vacancy is placed is completely symmetrical. This fact confirms
the correctness of the wave function obtained in~\ref{Eq13}. The
wave function is non-zero on sites $d$, due to the presence of the
phase factor of $\exp(ik_x/2)$,  but its magnitude is symmetrical.
The local density of states is zero on all atoms of type $a$ located
on the armchair rows, which confirms the condition~\ref{Eq11}. The
characteristics outlined in the graph align perfectly with the
predictions of the proposed model confirming qualitatively its
validity. The figure also shows an interesting outcome: much like
the wave function corresponding to the phosphorene's localized
state, the wave function associated to the present state is also
strongly localized. So, it seems that if two or more vacancies are
located next to each other at a distance of more than a few atoms,
due to the non-overlapping of their wave functions, they do not
interact. The following section will examine this issue and
demonstrate that, when there are two vacancies situated at a
distance from each other in the zigzag orientation, the localized
states in Haldane's model can interact with each other even over
substantial distances. It is interesting how this statement does not
hold true for two vacancies positioned in the armchair direction,
where the interaction between the two localized states is absolutely
absent.\par
\begin{figure}[t]
\begin{center}
\includegraphics[scale=0.35]{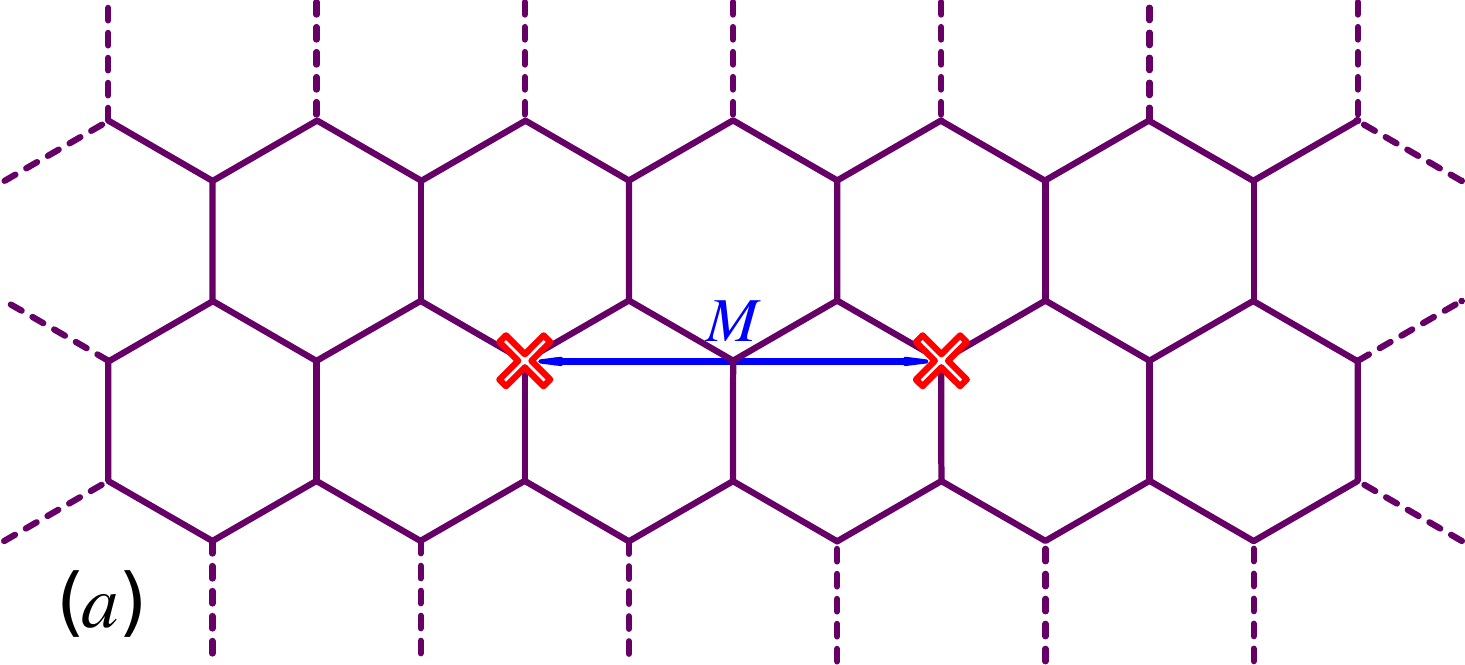}
\includegraphics[scale= 0.7]{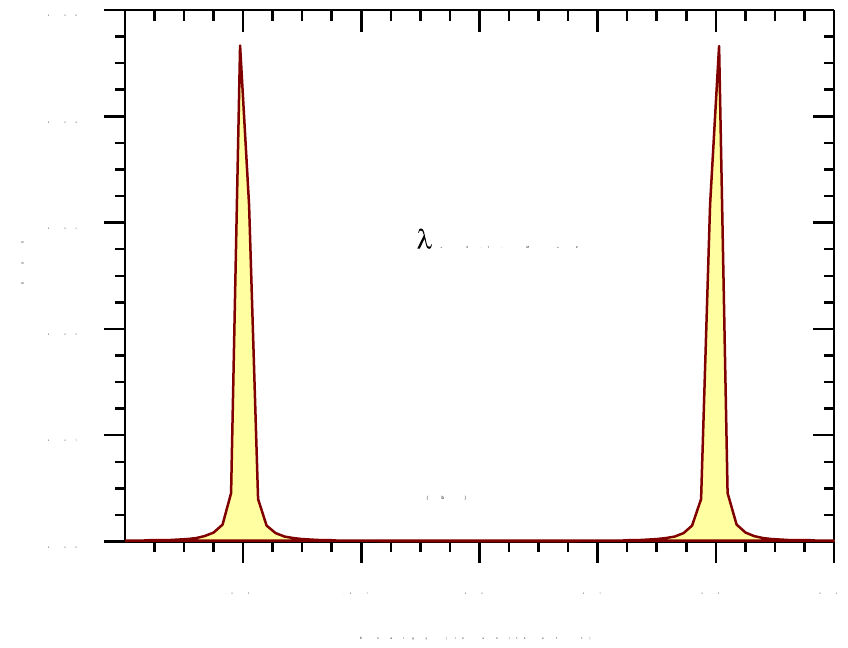}

\end{center}
 \caption{$(a)$ Two vacant sites created on a zigzag row of
a honeycomb sheet at a distance of $M$ zigzag chains, $(b)$ Density
of the states~(DOS) for the Haldane model corresponding to the
configuration of~$(a)$ in terms of energy for $\lambda = 0.1$ and
$M=2$. As is seen, for two energies of $\epsilon = \mp 2t$, DOS is
non-zero in the gap area. \label{Fig04}}
\end{figure}
\section{The hopping between two localized states and the impurity band\label{Sec03}}
Within this section, we demonstrate that the presence of two
vacancies on two sites along a zigzag  chain of a Haldane honeycomb
lattice leads to the generation of a hopping interaction term
between the two created localized states around these vacancies. As
a result, two vacant sites will function similarly to a two-level
quantum dot. We will show that if two vacancies are arranged in a
zigzag direction, the hopping interaction between these states is
significant even at long distances between the vacancies. But, If
the vacancies are placed in an armature row, no hopping interaction
will occur between them.\par
The schematic representation in part~(a) of figure~\ref{Fig04}
illustrates the issue examined in this section. The figure indicates
that two vacancies have been assumed to create on two lattice sites
in the zigzag direction, with the distance between them being equal
to $M$ zigzag chains. For the introduced configuration with $M = 2$,
the energy local density of states~(DOS) is plotted in terms of
energy in part~$(b)$ of the figure. Evidently, even though each
individual localized state possesses zero energy, the energy becomes
non-zero when these states are in proximity to each other. This fact
indicates that when the localized states are arranged adjacent to
each other in a zigzag row, they will interact with each other. By
representing the states around the vacancies as $|\psi_L^1\rangle$
and $|\psi_L^2\rangle$, in terms of the mentioned feature we can
express
\begin{equation}\label{Eq15}
H = \epsilon \left|\psi_1^L\right\rangle \left\langle\psi_2^L\right|
+ h.c.,
\end{equation}
as the Hamiltonian describing the tunneling between the zero-energy
localized modes around the vacant sites. In this Hamiltonian in
which $h.c.$ stands for Hermitian conjugate and $\epsilon$ refers to
the hopping energy of the zero-energy localized modes surrounding
the vacancies. This form of the Hamiltonian explicitly implies that,
the energy eigenstates are two symmetric and antisymmetric
combination of the localized states $\left|\psi_L^1\right\rangle$
and $\left|\psi_L^2\right\rangle$, whose eigenenergies are equal to
$+\epsilon$ and $-\epsilon$, respectively.\par
It is easy to examine the above issue  for two on-site adjacent
vacancies created in the armchair direction. Our examination
indicates that, in this case, there is no interaction between the
localized states and as a result the hopping interaction energy
between the created localized states is zero. The non-overlapping of
the zero-energy localized modes in the armchair direction causes the
absence of any interaction energy between the states in this
direction.\par
\begin{figure}[t]
\begin{center}
\includegraphics[scale=1]{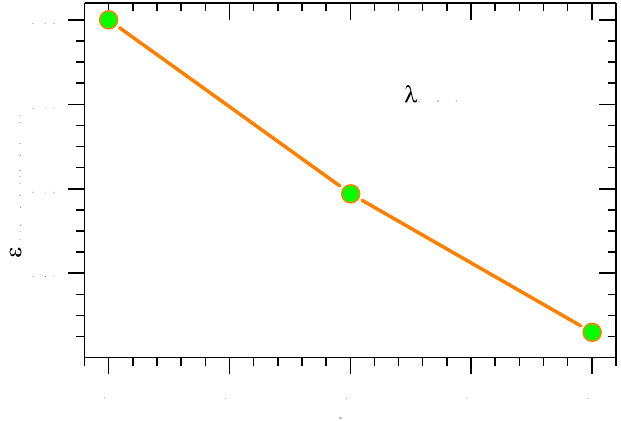}

\end{center}
\vspace*{-3mm} \caption{The effective interaction~(hopping) energy
between the zero-energy localized states around two on-site
vacancies in a zigzag row of the Haladane lattice is plotted as a
function of the distance between the vacant sites. \label{Fig05}}
\end{figure}

Another issue we would like to examine is the distance dependence of
the tunneling strength between the localized states of
$|\psi_1^L\rangle$ and $|\psi_1^L\rangle$. Is it possible to tunnel
between these states when the distance between the vacant sides is
considerable? The Hamiltonian responsible for tunneling is given in
bases $|\psi_1^L\rangle$ and $|\psi_1^L\rangle$ in
equation~\ref{Eq15}. As is seen, the strength of tunneling between
the specified mods depends on the interaction~(hopping) energy of
$\epsilon$. The larger the $\epsilon$, the greater tunneling
strength. Tunneling between the states at long distances requires
that $\epsilon$ be not zero at such distances. For the configuration
presented in panal~(a) of figuer~\ref{Fig04}, the effective
interaction~(hopping) energy, $\epsilon$, is calculated as a
function of the distance between the vacant sites, $M$. The result
is depicted in figure~\ref{Fig05} for $\lambda=0.1$. As is seen,
with an increase in $M$, there is an almost linear decrease in
$\epsilon$, but the calculated hopping energy remains significant
even over long distances. As $\lambda$ decreases, the localization
length of the edge states will increase, leading to higher tunneling
strength. The behavior exhibited in figures~\ref{Fig04}
and~\ref{Fig05} implies that two vacancies in Haldane's model  acts
similar to a two-level quantum dot. Also, it can be concluded that
if the on-site vacancies are periodically distributed in the zigzag
direction in a ribbon of the specified Haldane lattice, they behave
like a tight-binding model and we expect a sinusoidal dispersion to
be created in the center of the gap.\par
\begin{figure}[t]
\begin{center}
\includegraphics[scale=0.5]{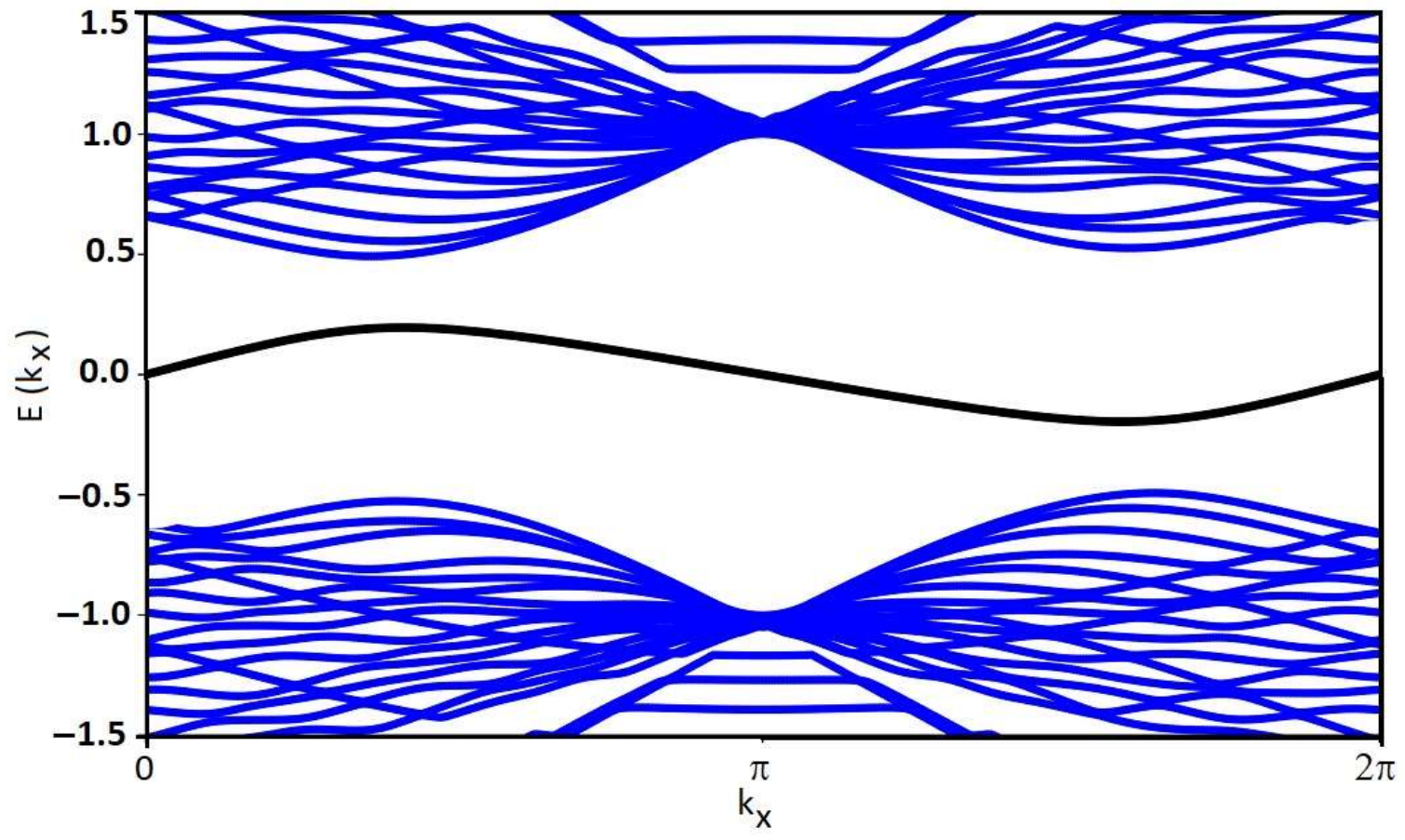}

\end{center}
\vspace*{-3mm} \caption{The band structure of a Haldane armchair
ribbon with $\lambda=0.1$, in which the impurities are regularly
distributed on a zigzag line with a distance of $M=2$, the
appearance of a sinusoidal impurity band in the gap region
evident.\label{Fig06}}
\end{figure}

As a final note, we are interested in how the energy band structure
of a finite-width ribbon is affected by the presence of the onsite
vacancies in the Haldane lattice of a TMD monolayer. To address this
issue, we consider an armchair side ribbon with finite width but an
effective infinite length. The (hopping) interaction energy for this
structure can be calculated by the results of equations~\ref{Eq13}
and~\ref{Eq15}. As it has been demonstrated in
references~\cite{Amini01} and~\cite{Kondo01}, the finite width of
the ribbons causes a small gap in the energy band of its topological
structure. This phenomenon, which is caused by the overlap of the
on-site zero-energy localized state such as $|\psi_1^L\rangle$ and
$|\psi_1^L\rangle$, is generally called the finite-width effect. As
has been explored in references~\cite{Sadeghizadeh01}
and~\cite{Kondo01}, the finite-width effect is for the localized
edge states in an armchair Haldane nano-ribbon is more considerable
for larger widths. In fact, it has been demonstrated that when the
width of the ribbon is comparable to the localization length of the
edge states, a minor gap due to the overlapping of the two edge
states appears in the ribbon's topological structure. In the
analysis of the formation of this gap, it can be shown that the
overlapping of $|\psi_1^L\rangle$ and $|\psi_1^L\rangle$ creates an
effective interaction between these states. According this property,
it is presumed that the vacancies are arranged along a zigzag row of
an armchair ribbon with regular and repeated distances, and the
electron band structure of Haldane's model is computed using the
obtained zero-energy localized states given equation~\ref{Eq13}. As
expected, each of the examined arrangements for this periodic
vacancy pattern display a sinusoidal dispersion at the center of the
energy gap. As an example, the band structure of an armchair ribbon
of the Haldane model with $\lambda=0.1$, in the presence of a
periodic line of vacancies with a distance of $M=$ on a zigzag
chain, is shown in figure~\ref{Fig06}. As is seen, the formation of
an impurity band in the energy gap of the Haldane model is evident.
This impurity band exhibits a sinusoidal behavior and is applicable
to produce the phenomena such as the topological Fano
effect~\cite{Amini01,Zangeneh01,Ji01,Amini02}.
\section{Conclusion\label{Sec04}}
In this study, we have investigated the effects of vacant sites in
2D Haldane materials, focusing on the formation of localized states
around single and multiple vacancies. Evidence demonstrated the
existence of a zero-energy localized mode around each vacancy, which
can be constructed by combining the armchair edge states. By
utilizing appropriate Fourier and unitary transformations, we have
obtained the wave function of the created localized state around a
single vacancy in the Haldane lattice, and we have demonstrated the
formation of zero-energy localized modes similar to those of a
quantum dot. Furthermore, we have studied the hopping interaction
energy between localized modes formed around two vacant sites,
revealing notable interactions for vacancies situated in the zigzag
direction. Study has demonstrated that the coupling between two
zero-energy localized modes, which surround two vacant sites in a
zigzag row, exhibits characteristics similar to that of a two-state
quantum dot. Additionally, we have investigated the characteristics
of the impurity band created in the electronic band structure of the
system when introducing vacancies periodically on a lattice zigzag
chain. The impurity band that is produced bears a strong resemblance
to a one-dimensional tight-binding model. Our findings contribute to
a better understanding of the behavior of vacancies in the Haldane
model and their implications for the emergence of impurity bands.
The results of this study provide valuable insights into the unique
properties of Haldane materials in the presence of vacancies, with
potential implications for the design and study of novel quantum
materials.

\end{document}